\documentclass[useAMS]{mn2e}
\usepackage{amsmath}
\usepackage{textcomp}
\usepackage{amssymb}
\usepackage[pdftex]{graphicx}
\usepackage{amssymb}
\usepackage{bm}

\usepackage[margin=.8in, paperwidth=8.5in, paperheight=11in]{geometry}
\usepackage{multicol}
\usepackage{epsfig}
\usepackage{graphicx}

\begin{document}

\title{Rossby Vortices in Thin Magnetised Accretion Discs}
\author[L. Matilsky et al.]
{\parbox{\textwidth}{L. Matilsky$^{1,2}$, S.~Dyda$^{1,3}$,
R.V.E.~Lovelace$^{4}$,
P.S. Lii$^{4}$}\vspace{0.3cm}\\
 $^{1}$Department of Physics, Cornell University, Ithaca, NY 14853, USA; loren.matilsky@colorado.edu\\
 $^{2}$JILA \& Department of Astrophysical and Planetary Sciences, University of Colorado, Boulder, CO 80309, USA\\
 $^{3}$Department of Physics \& Astronomy, University of Nevada, Las Vegas, NV 89154, USA\\
 $^{4}$Department of Astronomy, Cornell University, Ithaca, NY 14853, USA\\
}

\date{Accepted 2018 July 23. Received 2018 July 2; in original form 2017 November 23}
\pagerange{\pageref{firstpage}--\pageref{lastpage}}
\pubyear{2018}

\label{firstpage}
\maketitle

\begin{abstract}
We study the Rossby wave instability (RWI) in a thin accretion disc threaded by an initially toroidal magnetic field using the magnetohydrodynamics (MHD) code \textsc{PLUTO}. We find that for plasma $\beta$ in the range $10^1 \leq \beta \leq 10^3$, the growth rate and late-time density are suppressed, whereas for plasma $\beta > 10^3$, the magnetic field has negligible effect. The initially toroidal field is twisted inside the vortex and expelled at late times. This creates a radially directed field outside the vortex region, which can be understood via a simple kinematic model of a magnetic field in a rotating fluid and may be observable via polarised dust emission.
\end{abstract}


\section{Introduction}
Vortex formation in accretion discs is important in the context of planet formation, as it provides a mechanism for concentrating dust grains. It may also help explain non-axisymmetric features observed in transitional discs by the Combined Array for Research in Millimeter Astronomy (CARMA; Isella et al. 2013) and the Atacama Large Millimeter Array (ALMA; Casassus et al. 2013; van der Marel et al. 2013). 

A possible mechanism for the formation of vortices in discs is the Rossby wave instability (RWI). The RWI operates near local peaks in the density or specific entropy in the disc. Such extrema could occur near the edges of dead zones or near gaps in the disc that are opened up by protoplanets (see Lovelace \& Romanova 2014 for a review). 

Theoretical studies of the RWI in axisymmetric thin discs were carried out in the linear regime (Lovelace et al. 
1999; Li et al. 2000) and in the nonlinear regime (Li et al. 2001). These studies showed that the instability growth rates $\gamma$ scale like $\gamma\sim 0.2 \Omega_K$, where $\Omega_K$ is the Keplerian angular velocity local to the vortex. These results were confirmed using numerical hydrodynamic (HD) simulations in 2D (de Val-Borro et al. 2007; Lyra et al. 2009a; Lyra et al. 2009b) and 3D (Varniere \& Tagger 2006; Lin 2014). Numerical bi-fluid models showed that dust in the disc is trapped by Rossby vortices, providing a possible mechanism for planetesimal formation (Meheut et al. 2012; Zhu et al. 2014; Zhu \& Baruteau 2016; Baruteau \& Zhu 2016). 
 
Observations suggest that accretion discs are weakly magnetised, necessitating study of the RWI in the presence of magnetic fields (Kylafis et al. 1980; Nagase 1989). Some progress has been made in this direction, primarily using linear theory. Yu \& Li (2009) performed a linear stability analysis of thin discs with a poloidal magnetic field and found the field to have an overall stabilizing effect on the RWI. Yu \& Lai (2013) have shown that the inclusion of a large-scale poloidal magnetic field can enhance the growth rate of the RWI by a factor of 2. Gholipour and Nejad-Ashgar (2015) performed the linear evolution of the RWI in a thin disc threaded by a toroidal field and found that the field damps perturbations in the disc. 3D MHD simulations by Flock et al. (2015) found that the RWI operates in magnetised accretion discs near the outer dead zone and that the resulting features should be observable with ALMA. 

In addition to altering the RWI, magnetic fields in vortices may also generate novel and observable physical effects. In magnetised discs, coupling between vortices and the magnetic field may trigger the accretion-ejection instability (Tagger \& Pellat 1999). This could explain the quasi-periodic oscillations (QPOs) of Sagitarius A* (Tagger \& Melia 2006) or microquasars in general (Tagger \& Varniere 2006). Dust grains may align with the magnetic field due to radiative torques (see Lazarian 2007 for a review), possibly allowing measurements of magnetic field morphology using polarised dust emission (Stephens et al. 2014).

We perform 2D MHD simulations of a weakly magnetised accretion disc using the \textsc{PLUTO} code (Mignone et al. 2007; Mignone 
et al. 2012). We set the initial 2D profiles of the fluid variables so that the disc begins in hydrostatic and thermal equilibrium and is axisymmetric, except for a perturbation in the density profile. PLUTO solves MHD equations for the evolution of the disc numerically and we study the behavior of the resulting Rossby vortices. We examine the dependence of the growth rate and saturation of the instability as a function of the disc's initial plasma $\beta$ (ratio of gas pressure to magnetic pressure) and study the evolving morphology of the magnetic field. 

In Section \ref{sec:theory}, we describe our numerical methods, including our initial and boundary conditions, as well as a summary of the PLUTO code's formulation of the MHD equations. In Section \ref{sec:field_twisting}, we describe a kinematic model for the twisting of magnetic field lines in a vortex. In Section \ref{sec:results}, we present the results of the simulations. We measure the growth rates of the vortices and their saturation values. We study the evolution of the field inside the vortex and along its boundary. We show that magnetic field lines become twisted in the vortex region and, at late times, expelled. In Section \ref{sec:discussion}, we explore some possible observational consequences of our work and outline future directions of inquiry. 

\section{Numerical Setup}
\label{sec:theory}
We perform simulations with the \textsc{PLUTO} code (version 4.1, Mignone et al. 2007; Mignone et al. 2012) in 2D polar coordinates $(R,\phi)$ using the MHD physics module. We use PLUTO to numerically solve the ideal, inviscid and resistance-less MHD equations,
\begin{subequations}
 \begin{align}
\frac{\partial \Sigma}{\partial t} &=  -\nabla \cdot \left( \Sigma \bm{v} \right), \\
 \frac{\partial (\Sigma \bm{v})}{\partial t} &= - \nabla \cdot \bm{\mathcal{T}} -\Sigma \nabla \Phi, \\
\frac{\partial \bm{B}}{\partial t}  &= \nabla \times (\bm{v} \times \bm{B}) \label{eq:induction} \\
\text{and}\ \ \ \ \ \frac{\partial P}{\partial t} &= -\nabla \cdot \left( P \bm{v} \right).
 \end{align}
\end{subequations}
Here, $\Sigma$ and $P$ are the vertically integrated surface density and pressure, respectively, $\bm{v} = v_R\hat{\bm{e}}_R+v_{\phi}\hat{\bm{e}}_\phi$ is the 2D velocity, $\bm{B} = B_R\hat{\bm{e}}_R+B_{\phi}\hat{\bm{e}}_\phi$ is the 2D magnetic field and $\Phi(R) = -GM_*/R$ is the gravitational potential outside a star of mass $M_*$, with $G$ the universal gravitational constant. We ignore the self-gravity of the disc. The unit vectors $\hat{\bm{e}}_R$ and $\hat{\bm{e}}_\phi$ refer to the standard radial and azimuthal directions (respectively) in polar coordinates. The components of the momentum flux density tensor $\mathcal{\bm{T}}$ are given by
\begin{equation}
 \mathcal{T}_{ik} =P\delta_{ik}+ \Sigma v_i v_k + \left( \frac{\bm{B}^2}{8 \pi} \delta_{ik} - \frac{B_i B_k}{4 \pi} \right),
 \label{eq:mom_flux_tensor}
\end{equation}
where the term in parentheses, the so-called Maxwell stress, is responsible for the Lorentz force. In equation \eqref{eq:mom_flux_tensor}, $i$ and $k$ range over the two polar coordinates $R$ and $\phi$, while $\delta_{ij}$ is the Kronecker delta, equal to 1 for $i=j$ and 0 for $i\neq j$. We use an ideal equation of state so that $S \equiv P/\Sigma^{\Gamma}$ is a conserved quantity for the fluid in the Lagrangian sense. We take $\Gamma = 5/3,$ corresponding to an ideal gas with three translational degrees of freedom. 

We use PLUTO to solve the preceding equations in dimensionless form. The adiabatic (M)HD equations are scale-invariant, so the simulations technically represent a wide range of physical systems with the same non-dimensional parameters. However, it is helpful to use a real observational disc for comparison to our ``code units." We choose the example of the gas-rich disc surrounding the binary system V4046 Sgr as characterised by Rosenfeld et al. (2012). To compare with this system, we choose as a unit of length $l_0=R_0=25$ AU to represent the initial location of the density perturbation (this is chosen to agree with the overall extent of 45 AU reported for V4046 Sgr). We choose our unit of time to be $t_0=(GM_*/R_0^3)^{1/2}\equiv P_0/2\pi$, the Keplerian rotation period $P_0$ at the location of the perturbation divided by $2\pi$ (with the total mass of V4046 Sgr being $M_*=1.75 M_\odot$, this yields $t_0 \approx15$ yr). We also define the Keplerian rotation frequency at the location of the perturbation as $\Omega_0\equiv2\pi/P_0$ ($=1$ in our units). Finally, we set our unit of mass by assuming that $\Sigma_0$ (the amplitude of the surface density of the unperturbed disc at $R_0$) is equal to $30\ \rm{g}/\rm{cm}^2$, yielding $m_0=\Sigma_0l_0^2\approx2.2M_J$, where $M_J=1.90\times10^{30}$ g is the Jovian mass. This choice for $\Sigma_0$ is consistent with the disc in V4046 Sgr when we assume a CO abundance of $10^{-4}$ as is typical of dark molecular clouds. Derivative units are velocity $v_0=l_0/t_0\approx7.9\ \rm{km}/\rm{s}$, density $\rho_0=m_0/l_0^3\approx8.0\times10^{-14}\ \rm{g}/\rm{cm}^3$ and magnetic field $b_0 = \sqrt{4\pi\rho_0l_0v_0^2} \approx 0.79$ G.

The key criteria for growth via the RWI is an extremum in the quantity
\begin{align}
\mathcal{L} = \frac{\Sigma \Omega}{\kappa^2} \left( P \Sigma^{- \Gamma}\right)^{2/\Gamma}, 
\end{align}
where $\Omega = v_{\phi}/R$ is the local angular rotation frequency of the disc and $\kappa^2 = 4 \Omega^2 + 2R\Omega (\partial\Omega/\partial R)$ is the squared epicyclic frequency (Lovelace et al. 1999). 

\begin{figure}
                \centering
                \includegraphics[width=0.5\textwidth]{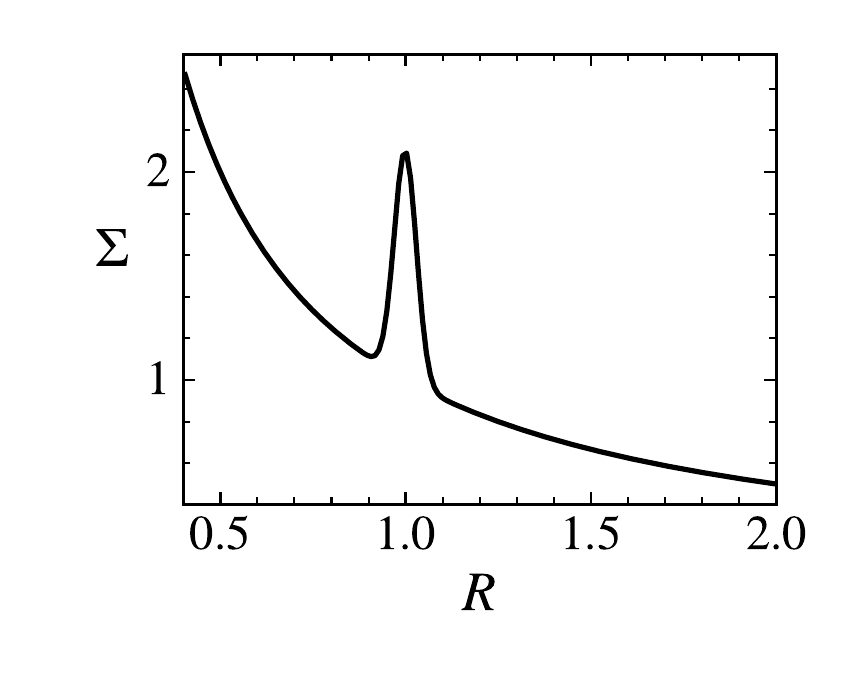}
                \includegraphics[width=0.5\textwidth]{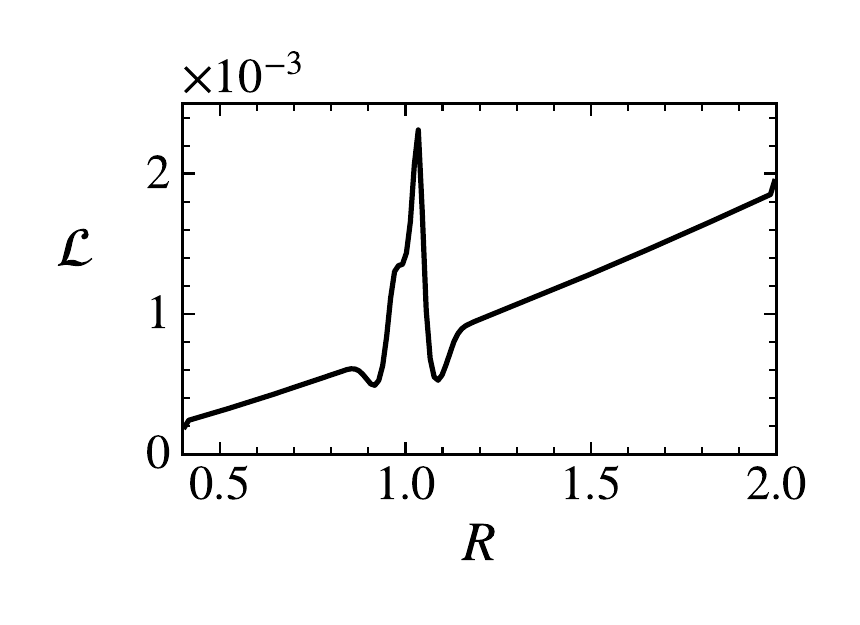}
        \caption{\textit{Top}: Initial, zero-longitude surface density profile $\Sigma(R,0,0)$, showing a local maximum centred around $R = R_0 = 1$. \textit{Bottom}: Initial, zero-longitude profile $\mathcal{L}(R,0,0)$, also showing a local maximum centred around $R  = R_0$.}
\label{fig:sigma}
\end{figure}

We implement the extremum in $\mathcal{L}$ via the initial surface density profile
\begin{align}
\Sigma_i(R,\phi) =
\bar{\Sigma} \Bigg{\{} 1 + c_1 e^{ -\frac{1}{2} \left(\frac{R - R_0}{\Delta R}\right)^2 }\Bigg{[} 1 + c_2 \
\cos(m\phi)\Bigg{]}\Bigg{\}},
\end{align}
where
\begin{align}
 \bar{\Sigma} = \frac{\Sigma_0}{R}
\end{align}
is the equilibrium density distribution for 2D discs and $\Sigma_0 = 1$ is the surface density at $R_0$. 
The parameter $c_1 = 1$ is the amplitude of the density perturbation that leads to formation 
of the RWI. The perturbation is centred around the reference radius $R_0$, while $\Delta R = 0.03$ controls the perturbation's width. The parameter $c_2 = 0.01 $ is the 
amplitude of a perturbation in the azimuthal direction, which is introduced to induce the formation of Rossby vortices with azimuthal mode number $m = 3$. In Fig. \ref{fig:sigma}, we plot the initial, zero-longitude surface density $\Sigma(R,0,0)$ (top panel) and $\mathcal{L}(R,0,0)$ (bottom panel), which show initial ``bumps'' in their respective profiles. Note that the perturbation $c_1$ is not small, so we are exploring the Rossby instability well outside the linear regime. 

We choose the initial disc to be isothermal with a constant isothermal sound speed $c_s\equiv0.05$. For the physical disc mentioned above, this corresponds to an initial temperature of $\sim$$45$ K, which is slightly colder than the temperature of $\sim$$180$ K at $R=25$ AU reported by Rosenfeld et al. (2012), but still physically reasonable. The initial pressure profile is thus given by 

\begin{align}
P_i(R) = c_s^2 \Sigma_i(R).
\end{align}

The initial azimuthal velocity is set by requiring that the disc is in dynamical equilibrium in the radial direction. Ignoring the contribution from magnetic pressure, we choose
\begin{align}
v^2_{\phi, i} =
R \left(\frac{\partial\Phi}{\partial R} + \frac{1}{\Sigma_i}\frac{\partial P_i}{\partial R}\right)
\end{align}
and set the radial velocity $v_{R,i} = 0$.  

The initial magnetic field is taken to be purely toroidal ($B_R = 0$), with
\begin{equation}
B_{\phi,i} = \frac{B_0}{R}.
\end{equation}
We take field amplitudes $B_0$ in the range $1.780\times10^{-4} \leq B_0 \leq 3.160\times10^{-2}$, corresponding to a plasma $\beta = 2 P/|\bm{B}|^2$ in the range $5 \leq \beta \leq 5\times10^5$ at $R_0$. Though $\beta$ is a position-dependent quantity, when describing results, we will refer to our simulations by their initial $\beta$ values at $R = R_0$ and $\phi=0$. For comparison with previous HD studies, we examine a MHD simulation with $B_0 = 0$ as well. The initial field amplitude $B_0$ is an input parameter; our choices for $B_0$ yield $\beta$ values roughly evenly distributed in log-space within the range specified above.

We use a grid of size $(N_R,\ N_{\phi})$ = (150, 256), uniformly distributed in the region $0.4 \leq R \leq 2$ and $0 \leq \phi < 2 \pi$. We use periodic boundary conditions in $\phi$ ($\psi(R,0,t) = \psi(R,2\pi,t)$, where $\psi$ is any one of the fluid variables) and outflow (Neumann) conditions in $R$ (i.e., $\partial \psi/\partial R=0$ at the inner and outer boundaries of the disc).  We implement wave killing zones (see for example Fromang, Terquem \& Nelson 2005) in two regions at the inner ($0.2 \leq R \leq 0.4$) and outer ($2.0 \leq R \leq 2.2$) boundaries, which are necessary to prolong the simulations and study vortex evolution.

\section{Magnetic Field Twisting}
\label{sec:field_twisting}

\begin{figure}
                \centering
                \includegraphics[width=0.45\textwidth]{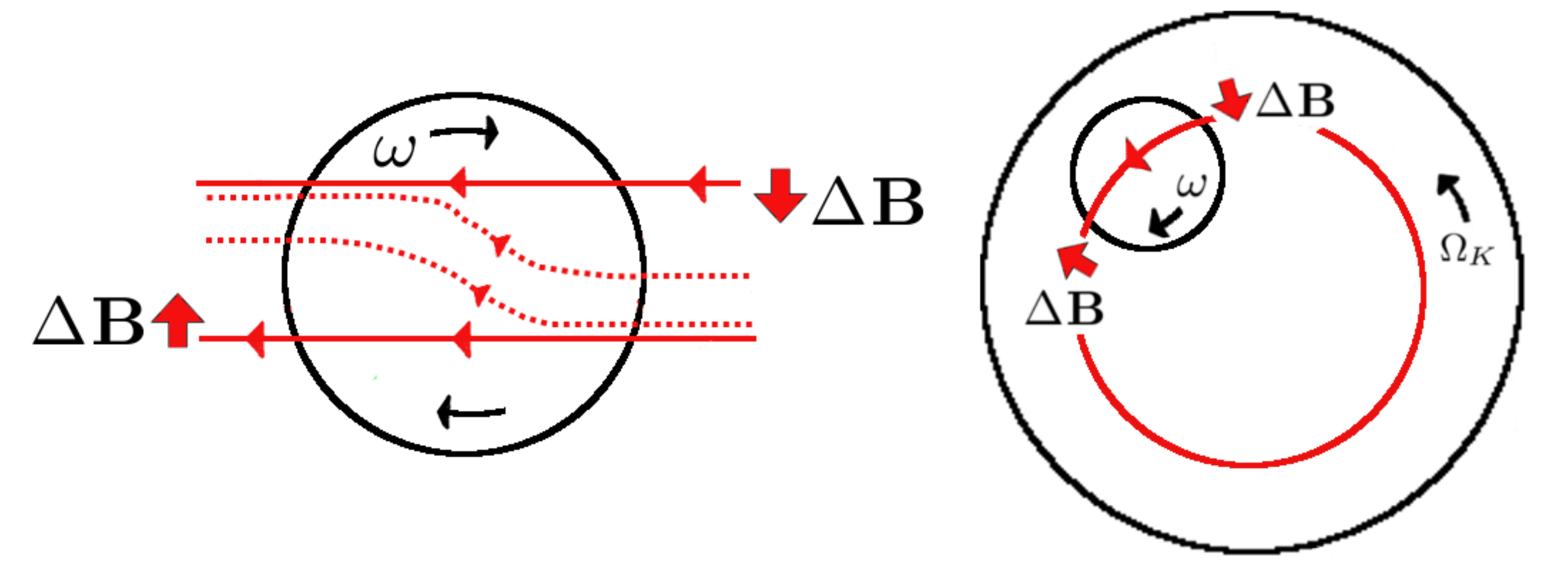}
        \caption{\textit{Left}: Sketch of the Moffatt (1983) model of a uniform field in a differentially rotating fluid. Initial magnetic field lines (solid red) become twisted (dotted red) at late times, causing an increase in the magnetic field in the direction indicated by the arrows. \textit{Right}: Treating Rossby vortices as in the Moffatt model implies that  there will be an increase in the $B_R$ field along the leading edge of the vortex and a decrease in $B_R$ near the trailing edge.}
\label{fig:cartoon}
\end{figure}

To develop an intuition for the dynamics of a toroidal field interacting with a vortex, we study a kinematic model developed by Moffatt (see Section 3.8 of Moffatt (1983) in particular). In what follows, we use lowercase polar ``vortex" coordinates ($r,\theta$) (and the corresponding Cartesian coordinates $(x,y) = (r\cos{\theta}, r\sin{\theta})$ to differentiate between our kinematic model of magnetic field twisting that is local to the vortex and the global evolution of the disc as traced by the coordinates $(R,\phi)$. 

Consider a uniform magnetic field initially pointing in the $-\hat{\bm{e}}_x$ direction: $\bm{B}(r,\theta,0) = -B_0 \hat{\bm{e}}_x$, coupled to a differentially rotating plasma with angular velocity $\omega(r)$ and turbulent magnetic diffusivity $\eta_t$. Although we do not implement a magnetic diffusivity explicitly using PLUTO, our simulations are fully non-linear and we find that the kinematics of the vortex are well described by invoking a turbulent diffusivity. By identifying the vortex directions $\hat{\bm{e}}_x$ and $\hat{\bm{e}}_y$ with the disc directions $-\hat{\bm{e}}_\phi$ and $\hat{\bm{e}}_R$ (respectively) at the location of the vortex, we can use the Moffatt model to represent our simulations' vortices threaded by an initially toroidal magnetic field. Since our vortices are anticyclonic, we  have $\omega(r) < 0$. A sketch of this model is shown in Fig. \ref{fig:cartoon} (left panel). Under the approximation of turbulent diffusion, the evolution of the field is given by
\begin{equation}
\frac{\partial \bm{B}}{\partial t} = \nabla \times \left( \bm{v} \times \bm{B} \right) + \eta_t \nabla^2 \bm{B}. 
\label{eq:turb_induction}
\end{equation}
With $\bm{v} =r\omega(r)\hat{\bm{e}}_\theta$, $\bm{B} = \nabla \times \bm{A}$ and the vector potential $\bm{A} = A \hat{\bm{e}}_z$, this can be written as
\begin{equation}
\frac{\partial A}{\partial t} + \omega (r)\frac{\partial A}{\partial \theta} = \eta_t \nabla^2 A,
\label{eq:twisting}
\end{equation}
with the initial condition $A(r,\theta,0) = -B_0 r \sin\theta$. 

\subsection{Early Times}
At early times, the perturbations in the fluid variables are small and the non-linear terms giving rise to the turbulent diffusion are negligible. Thus we set the RHS of equation (\ref{eq:twisting}) to zero and find the approximate solution 
\begin{align}
A(r,\theta,t)\approx -B_0 r \sin \left( \theta - \omega(r) t\right).
\label{eq:A_approx}
\end{align}
The magnetic field components are therefore
\begin{subequations}
	\begin{align}
	B_r &= \frac{1}{r}\frac{\partial A}{\partial\theta} = -B_0 \cos \left( \theta- \omega(r) t\right) \\
	\text{and}  \ \ \ \ \  B_{\theta} &= -\frac{\partial A}{\partial r} = B_0 \sin \left( \theta - \omega(r) t\right) \nonumber\\
	&-B_0 r \frac{d \omega}{d r} t \cos \left( \theta - \omega(r) t\right).
	\end{align}
\end{subequations}
We note that in the case of a non-differentially rotating disc, the second term on the RHS is zero and the field simply rotates at constant angular frequency $\omega$. 
This regime will eventually be suppressed, namely when the nonlinear pieces of the advection term $\nabla \times(\bm{v}\times\bm{B})$ on the RHS of equation \eqref{eq:induction} create significant turbulent diffusion. Using the first approximation \eqref{eq:A_approx}, we estimate 
\begin{align}
\eta_t\nabla^2A &\approx \eta_t\Bigg{[} \frac{1}{r}\frac{\partial}{\partial r}\Bigg{(}r\frac{\partial A}{\partial r}\Bigg{)} + \frac{1}{r^2}\frac{\partial^2A}{\partial \theta^2}\Bigg{]}\nonumber\\
	&= \frac{\eta_t}{r^2}\frac{d}{dr}\Bigg{(}r^3\frac{d\omega}{dr}\Bigg{)} B_0 t\cos(\theta - \omega(r)t)\nonumber\\
	&+ \eta_t\Bigg{(}\frac{d\omega}{dr}\Bigg{)}^2B_0rt^2\sin(\theta - \omega(r)t).
	\label{eq:approx_A_diff}
\end{align}
 Meanwhile, 
 \begin{align}
 \omega\frac{\partial A}{\partial \theta} \approx -B_0 r\omega(r)\cos(\theta - \omega(r)t).
 \label{eq:approx_A_adv}
 \end{align}
 To compare the magnitudes of the diffusion term and advection term in equation \eqref{eq:turb_induction}, we assume that $\omega(r)$ is smooth and define
 \begin{align}
 \omega_0 = {\rm{max}}\ |\omega(r)|
 \end{align}
 and 
 \begin{align}
 r_0 = \omega_0/{\rm{max}}\ \Big{|}\frac{d\omega}{dr}\Big{|}.
 \end{align}
The turbulent diffusion becomes comparable to the advective term when either of the coefficients of the sine and cosine terms in equation \eqref{eq:approx_A_diff} become comparable to the coefficient $B_0r\omega(r)\sim B_0r_0\omega_0$ of the cosine term in equation \eqref{eq:approx_A_adv}. This occurs when either  
\begin{equation}\label{eq:diff_crit}
\omega_0 t \sim R_{\rm{m}} \hspace{1.4cm} \text{or} \hspace{1.4cm}\omega_0 t \sim R_{\rm{m}}^{1/2},
\end{equation}
where 
\begin{align}
R_{\rm{m}} = \omega_0 r_0^2/\eta_t 
\end{align}
is the appropriate magnetic Reynolds number of the vortex. For high magnetic Reynolds numbers $R_{\rm{m}}>1$, the first condition in equation \eqref{eq:diff_crit} will be achieved at an earlier time, while for low magnetic Reynolds numbers $R_{\rm{m}}<1$, the second condition will be achieved at an earlier time.

\subsection{Late Times}
Equation \eqref{eq:diff_crit} shows that at late times, the regime of simple rotation of the field gives way to a steady state dominated by turbulent diffusion. Setting $\partial A/\partial t = 0$ in equation \eqref{eq:twisting} and choosing a separable ansatz 
\begin{equation}
A(r,\theta, t\rightarrow\infty) = - B_0f(r)\sin\theta=\text{Im}[{ - B_0 f(r) e^{i \theta}}]
\label{A_ansatz}
\end{equation}
yields
\begin{equation}
i \frac{\omega(r)}{\eta_t} f(r) = \frac{1}{r}\frac{d}{dr} \left( r f^\prime\right) - \frac{f}{r^2}. 
\label{f_efunc}
\end{equation}
Here the primes refer to radial derivatives. As long as $\omega(r)\rightarrow0$ as $r\rightarrow\infty$, the boundary condition on $f(r)$ should yield an undisturbed field $\bm{B}=-B_0\hat{\bm{e}}_x$ far away from the vortex, i.e., 
\begin{align}
f(r)\rightarrow r\ \ \ \text{as}\ \ \ r\rightarrow\infty .
\label{f_bc}
\end{align}
We consider the case of rigid rotation, for which
\begin{equation}
\frac{\omega}{\eta_t} = \begin{cases} -k_0^2 & r < r_0 \\
0 & r> r_0, \end{cases} 
\end{equation}
where $k_0$ is a real and positive constant. Taking into account the far-field boundary condition \eqref{f_bc}, the solution to \eqref{f_efunc} is then
\begin{equation}
f(r) = \begin{cases} DJ_1(pr) & r < r_0  \\
r + \frac{C}{r} & r > r_0 \end{cases} 
\label{complex_f}
\end{equation}
where 
\begin{equation}
p \equiv \Bigg{(}\frac{1+i}{\sqrt{2}}\Bigg{)}k_0
\label{eq:p}
\end{equation}
and $J_m$ is the $m^{th}$-order Bessel function of the first kind. The integration constants
\begin{equation}
C = r_0 \frac{ 2J_1(pr_0) - p r_0 J_0(pr_0)}{pJ_0(pr_0)}
\label{eqC}
\end{equation}
and 
\begin{equation}
D = \frac{2}{pJ_0(pr_0)}
\label{eqD}
\end{equation}
are chosen by requiring that the eigenfunction $f(r)$ and its derivative $f^\prime(r)$ are continuous (yielding continuous $B_r$ and $B_\phi$) across $r = r_0$. 

\begin{figure}
	\centering
	\includegraphics[width=0.45\textwidth]{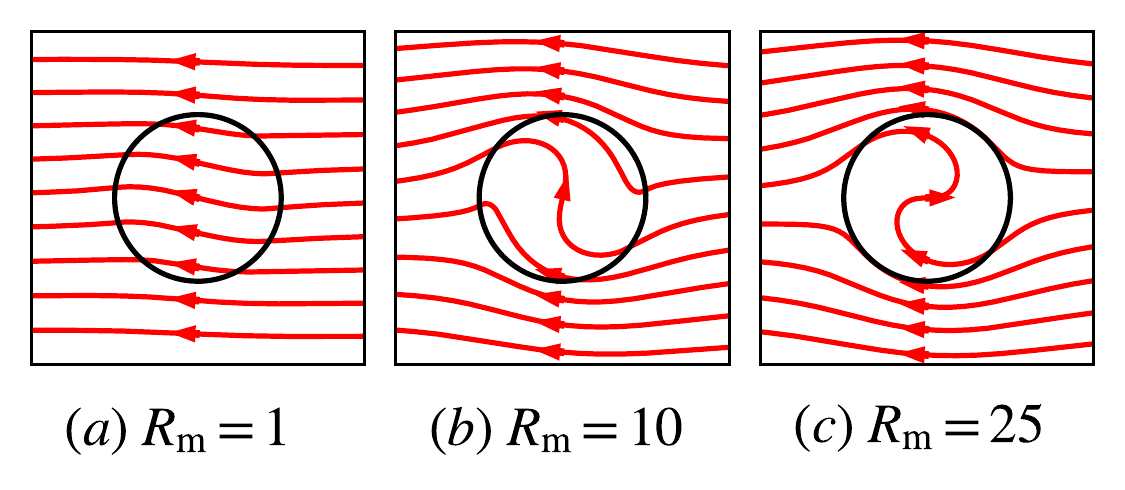}
	\caption{Late-time magnetic field streamlines for the Moffatt (1983) model of a uniform field in a rotating fluid for different values of the magnetic Reynolds number $R_{\rm{m}}$. The vortex region is marked with a black circle, with the angular velocity $\omega_0$ = constant within the vortex and zero outside. The sense of rotation is clockwise. (This figure closely resembles Fig. 3.3 in Moffatt (1983), although the field lines have been recomputed to match the orientation of our anticyclonic vortices in an initially toroidal magnetic field).}
	\label{fig:field_in_vortex}
\end{figure}

Fig. \ref{fig:field_in_vortex} shows the late-time orientation of the magnetic field for different regimes of turbulence indexed by the magnetic Reynolds number $R_{\rm{m}}$. The field lines are computed by evaluating evenly spaced contours of $A$ as defined by equations \eqref{A_ansatz}, \eqref{complex_f}, \eqref{eq:p}, \eqref{eqC} and \eqref{eqD}. In each panel, the black circle denotes the region of the vortex. For the weakly turbulent case (Fig. \ref{fig:field_in_vortex}a), the field is only slightly perturbed by the presence of the vortex. As the level of turbulence increases, however, the field becomes progressively more twisted in the direction of the vortex rotation, eventually reversing its direction entirely at vortex centre (Fig. \ref{fig:field_in_vortex}c). It should also be noted that fewer field lines exist within the vortex as the turbulence is increased, implying a weaker magnetic field magnitude overall within the vortex at late times.

\begin{figure}
                \centering
                \includegraphics[width=0.5\textwidth]{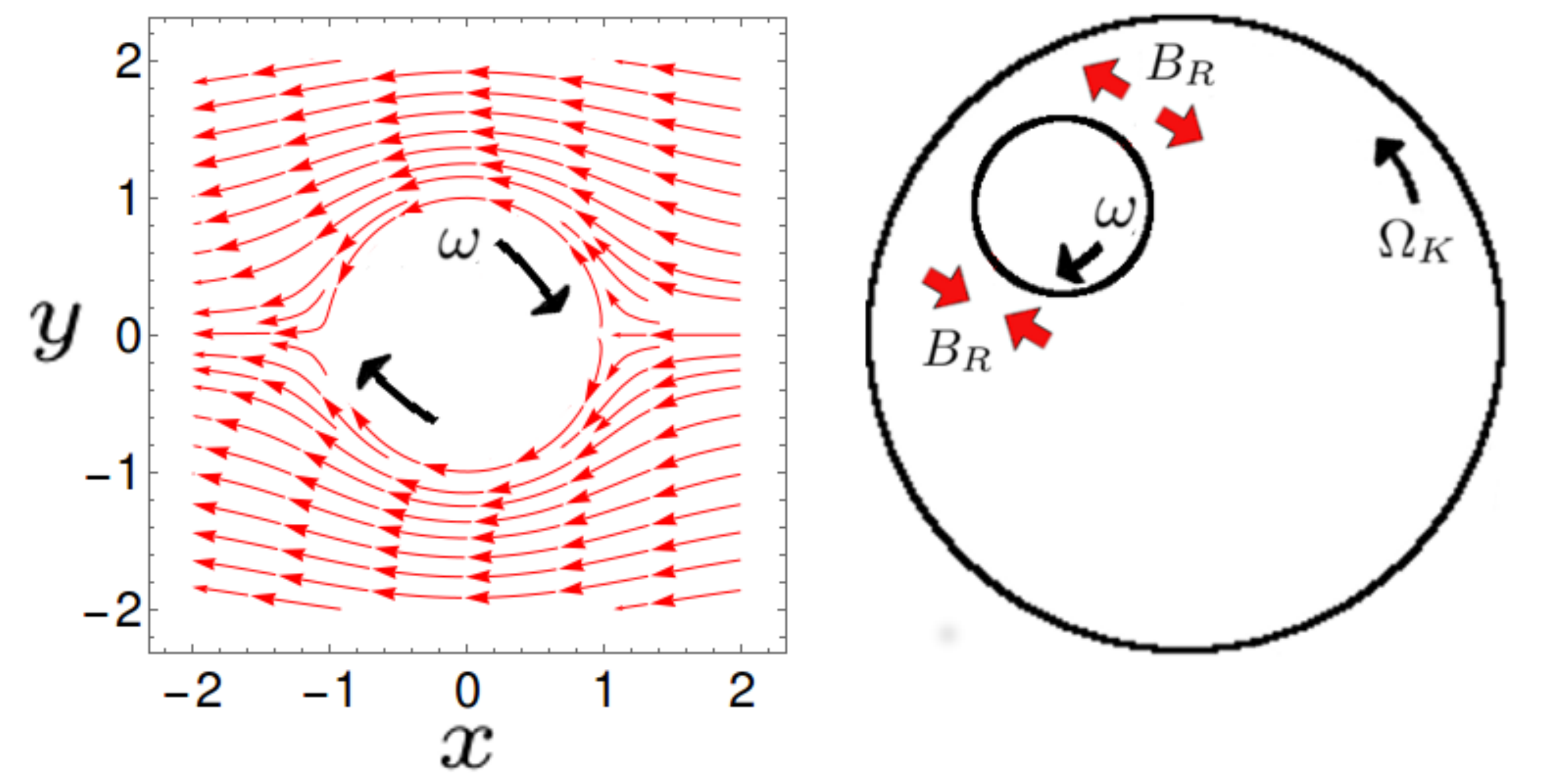}
        \caption{\textit{Left}: Late-time magnetic field streamlines for the Moffatt (1983) model of a uniform field in a rotating fluid in the limit of zero diffusivity ($R_{\rm{m}}\rightarrow\infty$).
        \textit{Right}: Treating Rossby vortices as in the Moffatt model implies the leading edge will act as a sink of radial magnetic field and the trailing edge as a source.}
\label{fig:late_time}
\end{figure}

In the limit $R_{\rm{m}} \rightarrow \infty$ (also $p$ and $k\rightarrow\infty$), equations \eqref{eqC} and \eqref{eqD} show that $C\rightarrow -r_0^2$, while $D\rightarrow0$. Thus, when there is no diffusion,
\begin{equation}
f(r) \rightarrow \begin{cases} 0 & r < r_0  \\
r - \frac{r_0^2}{r} & r > r_0 \end{cases} 
\label{f_nodiff}
\end{equation}
and 
\begin{equation}
A = \text{Im}[-B_0 f(r)e^{i\theta}] \rightarrow \begin{cases} 0 & r < r_0  \\
-B_0\Bigg{(} r - \frac{r_0^2}{r}\Bigg{)} \sin\theta & r > r_0. \end{cases} 
\label{A_nodiff}
\end{equation}
The fact that the vector potential vanishes inside the rotating region is akin to the skin effect in 
electromagnetism where currents, and therefore magnetic fields, are expelled from a rotating conductor. This is the first interesting effect expected for vortices at late times. 
The second is the orientation of the magnetic field. Outside the rotating region $r > r_0$, we compute the magnetic field components in the limit $R_{\rm{m}}\rightarrow\infty$,
\begin{align}
B_x &= \frac{\partial A}{\partial y} \rightarrow -B_0 + B_0\Bigg{(}\frac{r_0^2}{r^2}\Bigg{)}\cos{2\theta}\label{Bx_nodiff}\\
\text{and}\ \ \ \ \ B_y &= -\frac{\partial A}{\partial x} \rightarrow B_0\Bigg{(}\frac{r_0^2}{r^2}\Bigg{)}\sin{2\theta}.
\label{By_nodiff}
\end{align}
In Fig. \ref{fig:late_time} (left panel), we plot these magnetic field streamlines. The configuration is exactly analogous to the streamlines of an irrotational flow past a cylinder. In particular, we note that in the first and third quadrants, $B_y > 0$, while in the second and fourth quadrants, $B_y < 0$. For a toroidal field pointing in the direction of the rotation of the disc, we therefore expect the leading edge of the anticyclonic vortex to act as a sink of radial magnetic field and the trailing edge to act as a source of radial magnetic field (Fig. 
\ref{fig:late_time}, right panel). 

\section{Results}
\label{sec:results}

\begin{figure*}
                \centering
                \includegraphics[width=\textwidth]{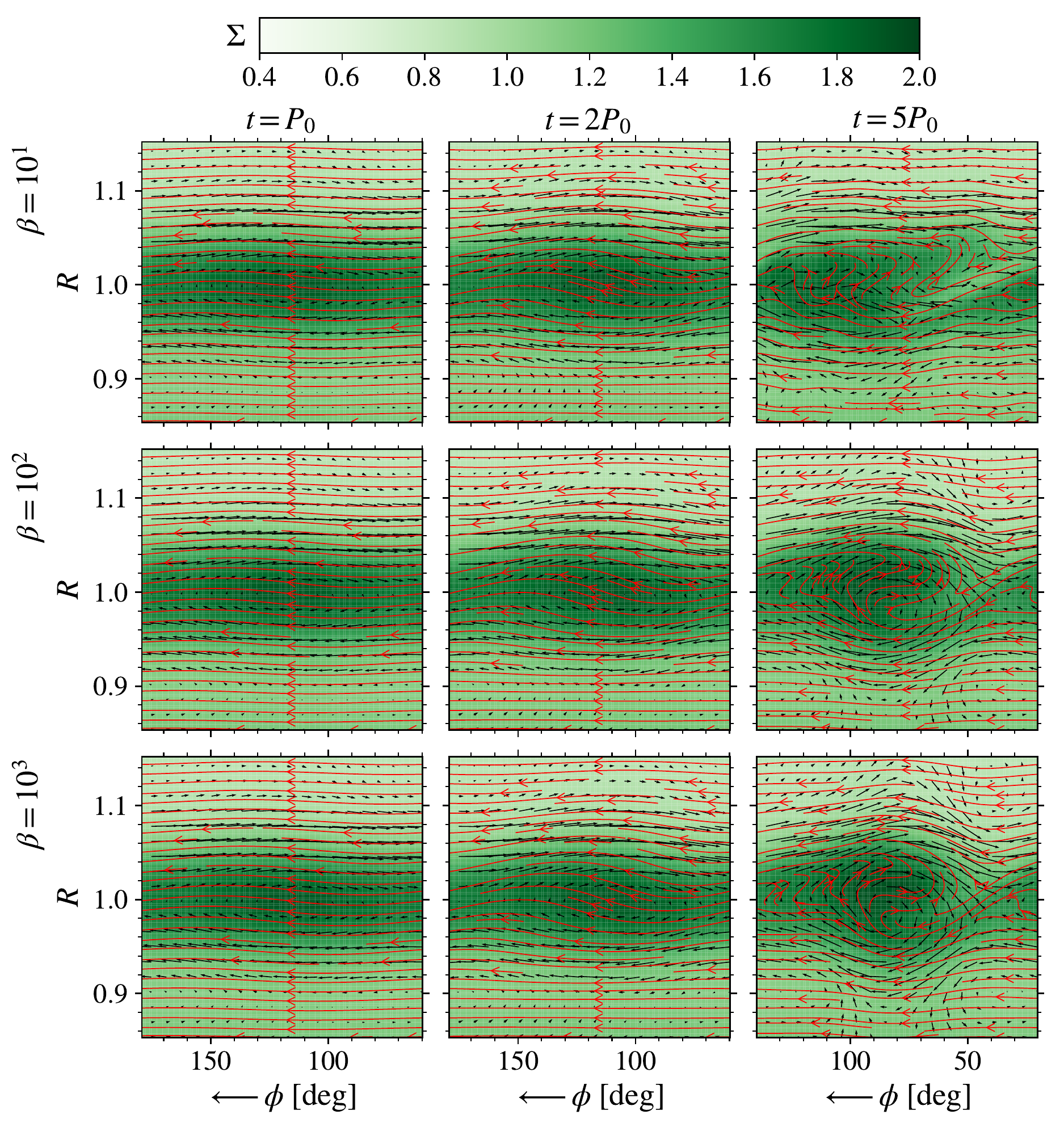}
        \caption{Plot of surface density $\Sigma$ (monochromatic green-scale), velocity field with the Keplerian rotation at $R = R_0 = 1$ subtracted out (black arrows) and magnetic field streamlines (red) for $10^{1} \leq \beta \leq 10^{3}$. In the weakest field cases (high $\beta$), the vortex grows quickly and has a well defined core with velocity streamlines wrapping around the density maximum. In cases with stronger magnetic field the vortex is not as well defined and takes longer to grow.}
\label{fig:streamline_summary}
\end{figure*}

In Fig. \ref{fig:streamline_summary}, we plot surface density $\Sigma$, the velocity fluctuation (full velocity field with the Keplerian rotation subtracted off) and the magnetic field streamlines at $t= P_0,\ 2P_0 \ {\rm{and}} \ 5P_0$ for plasma $\beta = 10^{1}, \ 10^{2}\ {\rm{and}} \ 10^{3}$. Subtracting off the Keplerian rotation from the velocity field allows us to explicitly see the heightened anticyclonic vorticity in the region of the density perturbation. In the weakest field cases ($\beta = 10^{2}$ and $\beta = 10^{3}$), the vortex grows quickly and has a well defined core with the velocity fluctuation wrapping around the density maximum. At early times, the magnetic field streamlines undergo distortions in the region of the vortex as seen in the kinematic model of Section \ref{sec:field_twisting}. At late times, the field lines become so twisted that their directions are reversed completely close to vortex center, reminiscent of the diffusion-dominated steady state depicted in Fig. \ref{fig:field_in_vortex}c. We thus estimate a magnetic Reynolds number of 
\begin{align}
R_{\rm{m}} \sim 25
\label{eq:Rm}
\end{align}
for the simulations in which the magnetic field reverses its direction at vortex centre. For lower values of the Reynolds number, the field would get twisted less (Fig. \ref{fig:field_in_vortex}a and \ref{fig:field_in_vortex}b), while for higher values, it would would have multiple twists inside the vortex. 

From the velocity fluctuation, we estimate the rotation rate of the vortex to be
\begin{align}
\omega_0\sim0.2,
\end{align}
while the radial extent of the vortex is
\begin{align}
r_0\sim0.1.
\end{align}
To achieve $R_{\rm{m}}=25$, we thus estimate the turbulent diffusivity empirically to be
\begin{align}
\eta_t = \frac{\omega_0r_0^2}{R_{\rm{m}}}\sim 10^{-4}.
\end{align}
In the standard prescription of Shakura \& Sunyaev (1973), the turbulent kinematic viscosity $\nu$ in accretion discs scales like
\begin{align}
\nu = \alpha \frac{c_s^2}{\Omega},
\end{align}
where $\alpha$ is a coefficient of order less than unity. For a unity magnetic Prandtl number ($\eta_t=\nu$), the effective $\alpha$ in our models due to turbulence is
\begin{align}
\alpha \sim \frac{\Omega_0\eta_t}{c_s^2} \sim 0.04.
\end{align}
While observations suggest $\alpha\sim0.1-0.4$, the value achieved in numerical simulations tends to be an order of magnitude smaller (King et al. 2007). Thus, we believe that the empirical value of $\eta_t\sim10^{-4}$ estimated for our simulations is reasonable. 

We see in the case of $\beta = 10^{1}$ that the density contrast of the vortex and surrounding disc is not as great as in the weaker-magnetisation cases. The vortex is also growing more slowly, with a less well defined velocity fluctuation wrapping the core at late times ($t = 5P_0$) compared to the weaker-magnetisation cases. Finally, the magnetic field lines in case $\beta=10^1$ are less distorted at late times, indicating a lower turbulent diffusivity.

\subsection{Vortex Evolution}  
Linear theory (Lovelace et al. 1999) predicts that the quantity $\Psi = \delta P/ \overline{\Sigma}$ grows exponentially in an unstable vortex. Here, $\delta P$ is the fluctuation of pressure in the disc with respect to its unperturbed value and $\overline{\Sigma}$ is the unperturbed value of the surface density. 

\subsubsection{Growth Rate at Early Times}
For each simulation, we define the key variable $\psi(t) = (P(t) - P_0)/\Sigma(t)$, where $P(t)$ and $\Sigma(t)$ are measured where the pressure is maximal in the vortex region and $P_0$ is the initial pressure at $R=1$ when the perturbation amplitude is set to zero. We adopt this definition (instead of the capital $\Psi$ from the linear theory of Lovelace et al. 1999) because our perturbation amplitude $c_1=1$ forces our simulations to be substantially outside the linear regime and we found the instability was more clearly visible in $\psi$ rather than $\Psi$. To estimate the growth rate of the instability, we calculate $\psi(t)$ for various magnetic field strengths and choose a minimum and maximum time ($t_{\rm{min}}=3$ and $t_{\rm{max}}=30$, respectively) to define the growth rate as
\begin{align}
\gamma \equiv \frac{\psi(t_{\rm{max}}) - \psi(t_{\rm{min}})}{t_{\rm{max}} - t_{\rm{min}}}.
\label{eq:rate_def}
\end{align}
The growth rate $\gamma$ may be thought of as the rate of fractional increase of $\psi$. In the linear regime of the instability, $\psi(t) - \psi(0)$ scales like $e^{\gamma t} - 1 \approx\gamma t$, making the definition of growth rate $\gamma$ in equation \eqref{eq:rate_def} appropriate. We choose a finite $t_{\rm{min}}=3$ (instead of $t_{\rm{min}}=0$) because of the initial decay of the fluid variables before the instability sets in.

\begin{figure}
                \centering
                \includegraphics[width=0.55\textwidth]{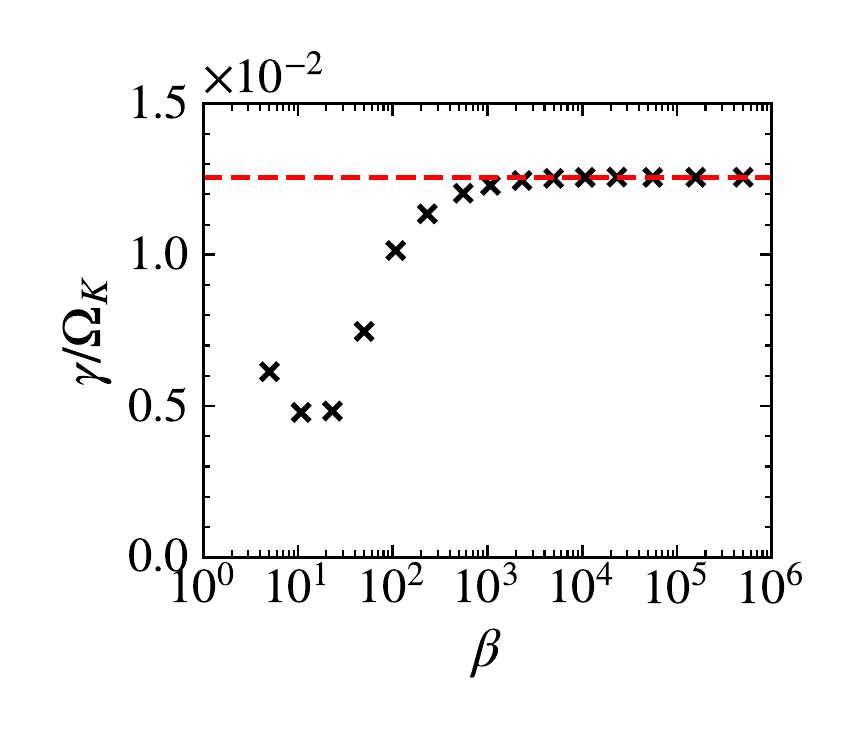}
        \caption{Growth rate $\gamma$ of the amplitude of the perturbation $\psi/\psi_0$ scaled to the local Keplerian angular frequency as a function of the initial local plasma $\beta$. For $\beta \gtrsim 10^3$, the growth rate converges to the value of the HD case ($\gamma/\Omega_K=1.26\times10^{-2}$) indicated by the dashed red line. For $\beta$ in roughly the range $10 \leq\beta \leq 200$, the growth rate has the power-law scaling $\gamma \sim \beta^{0.32} \ \Omega_K$.}
\label{fig:growth_rates}
\end{figure}
In Fig. \ref{fig:growth_rates}, we plot the growth rates scaled to the Keplerian orbital frequency at $R=1$ for different plasma $\beta$. For comparison, we include the growth rate 
of a disc simulated with PLUTO's purely hydrodynamic physics module.\footnote{For a sanity check, we note that the HD case yielded exactly the same results as the MHD case with $B_0=0$.} For plasma $\beta \gtrsim 10^3$, the growth rate asymptotically approaches the HD case. For $10^1 \lesssim \beta \lesssim 10^3$, the magnetic field suppresses vortex growth and the growth rate is small. We fit the dependence of growth rate on plasma $\beta$ to a power law and find that $\gamma \sim \beta^{0.32}  \Omega_K$. Extrapolating this power law to small values of $\beta$ we find that the vortex will not grow ($\gamma = 0$) for $\beta \ll 1$. We expect this since a magnetically dominated disc tends to suppress the small-scale shear associated with vortex growth. For $\beta = 5$ we see a slight deviation from the general trend with decreasing $\beta$. Yu and Lai (2013) see a similar effect in their linear analysis of a poloidal field threading a thin disc: growth rates decrease with increasing magnetisation until $\beta \sim 10$, after which they begin to increase. For such high disc magnetisations, vortex formation is highly suppressed in our simulations and it is difficult to determine an accurate growth rate from the definition in equation \eqref{eq:rate_def}  due to contamination from the initial decay of the fluid variables and the excitation of spiral density waves. We include the two points $\beta=5$ and $\beta=10$ in Fig. \ref{fig:growth_rates} for completeness, but because of our inability to run smaller-$\beta$ cases without the inner disc quickly becoming numerically unstable, we cannot say anything definitive about this regime.

\begin{figure}
                \centering
                \includegraphics[width=0.55\textwidth]{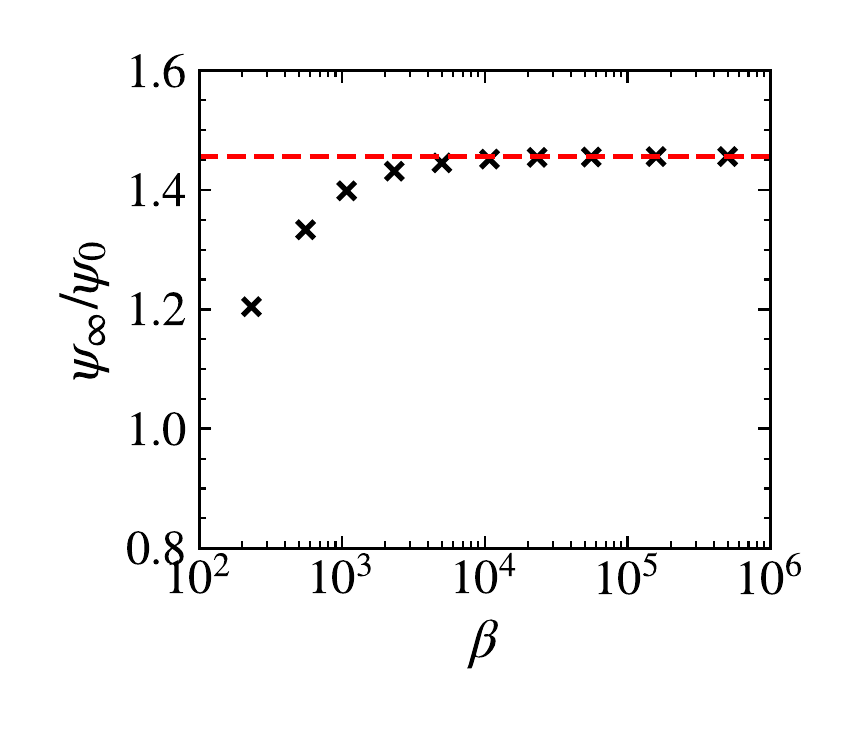}
        \caption{Late-time maximum amplitude of the perturbation $\psi_{\infty}/\psi_0$ as a function of the initial local plasma $\beta$. For $\beta \gtrsim 10^4$, saturation values converge towards the HD case ($\psi_\infty/\psi_0=1.46$) indicated by the dashed red line. For $\beta \lesssim 10^4$, $\psi_{\infty} \sim \beta^{0.075}\psi_0$.}
\label{fig:saturation_levels}
\end{figure}

\subsubsection{Maximum Amplitude at Late Times}
At late times ($t \gtrsim 5P_0$), vortex growth ceases and $\psi$ oscillates about a late-time value $\psi_{\infty}$. We estimate $\psi_{\infty}$ by averaging $\psi$ over $\sim$$10$ rotation period at late times. In Fig. \ref{fig:saturation_levels}, we plot $\psi_{\infty}/\psi_0$ as a function of $\beta$. For comparison, we include the saturation value of the HD case. At late times, simulations with stronger magnetisations tend to be less stable and it is more challenging to determine the asymptotic value $\psi_{\infty}$. For example, we see in Fig. \ref{fig:streamline_summary} that the strongest magnetisation case ($\beta = 10^1$) does not have as well defined vortex as the other cases do. Therefore, we only include cases with $\beta > 10^1$ for determining the maximum amplitude. For $\beta \gtrsim 10^4$, the late-time $\psi_{\infty}/\psi_0$ asymptotes to that of the HD case. For $10^3 < \beta < 10^4$, the late-time value decreases with plasma $\beta$. Fitting this regime to a power law yields $\psi_{\infty} \sim \beta^{0.075}$. Extrapolating the power law to high magnetisations, we find $\psi_{\infty} = \psi_0$ for $\beta \approx 15$. This is where the growth rate shown in Fig. \ref{fig:saturation_levels} experiences a small increase (with decreasing $\beta$) and we expect our model to break down as the disc transitions to a magnetically dominated regime.

We finally note that the level of turbulence in our simulations (and thus our computation of the dependence of instability parameters on plasma $\beta$) can potentially be affected by grid resolution. To verify that this is not the case, we calculate growth rates for our $\beta=10^2$ case run at three different resolutions: the original resolution ($N_R=150$, $N_\phi=256$), a coarser resolution ($N_R=75$, $N_\phi=128$) and a finer resolution ($N_R=300$, $N_\phi=512$). The growth rates $\gamma$ for the coarser and finer resolutions vary by no more than 4\% from that of the original resolution. The dependence of the instability's saturation value $\psi_\infty/\psi_0$ is a bit more complicated, since the higher the resolution, the faster the simulation becomes unstable. However, if we ignore the late-time oscillations in $\psi(t)$ associated with the beginnings of numerical instability (averaging the late-time value of $\psi(t)$ over $\sim$$3$ orbits instead of $\sim$$10$), we find that the saturation values in the coarser and finer resolutions vary by no more than 3\% from that of the original resolution.  We thus conclude that the numerical diffusivity plays a negligible role in the dependence of growth rate and saturation value on plasma $\beta$.

\subsection{Magnetic Field Evolution}
We compare the evolution of the field in the region of the vortex to the kinematic model described in Section \ref{sec:field_twisting}. We show plots for the lowest-magnetisation case $\beta = 5\times10^{5}$, in which the back reaction of the field on the vortex is least significant.  
\subsubsection{Early Times}
\begin{figure*}
                \centering
                \includegraphics{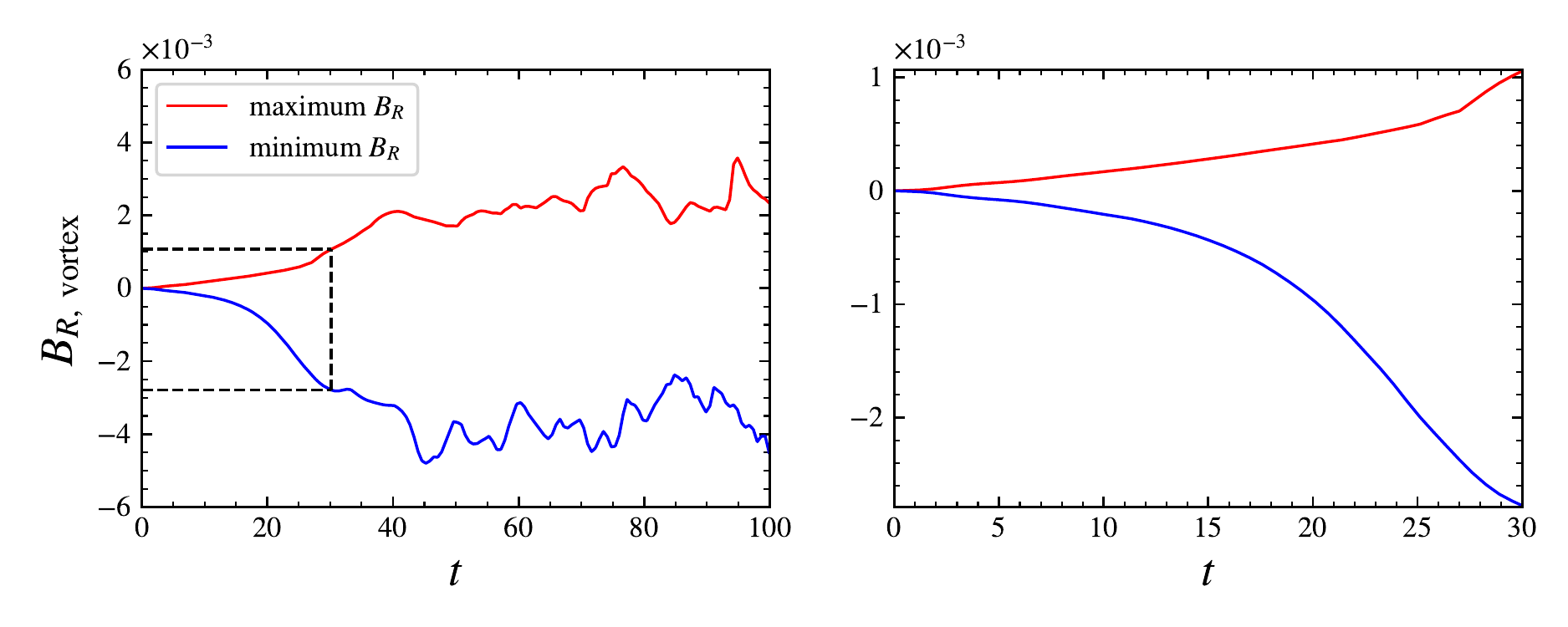}
        \caption{\textit{Left}: Maximum (red) and minimum (blue) $B_R$ in the vortex region for the case $\beta = 10^{2}$. The growth is roughly linear for the first few orbits of the vortex about disc centre ($t\lesssim15$) before entering a state with quasi-stationary maximum/minimum field. \textit{Right}: A blow-up of the region shown by the black, dashed box in the left panel, emphasizing the comparable linear growth rates of both positive and negative $B_R$ during early times.}
\label{fig:br_vs_time}
\end{figure*}
At early times, Moffatt's kinematic model predicts that the magnetic field is simply rotated by the vortex and so the radial component $B_R$ grows linearly for both positive and negative $B_R$.  In Fig. \ref{fig:br_vs_time}, we plot the maximum and minimum value of $B_R$ in the vortex as a function of time for $\beta=10^2$. For $t\lesssim15$, the positive and negative components grow linearly at the same rate, consistent with simple rotation of the field. It is during this time that plots of the field lines are slightly twisted in the vortex region (Fig. \ref{fig:cartoon}, left panel and Fig. \ref{fig:streamline_summary}, centre panels). After $t\sim15$, non-linear effects become important and turbulent diffusion comes to dominate the evolution.

\subsubsection{Late Times}

\begin{figure}
                \centering
                \includegraphics[width=0.45\textwidth]{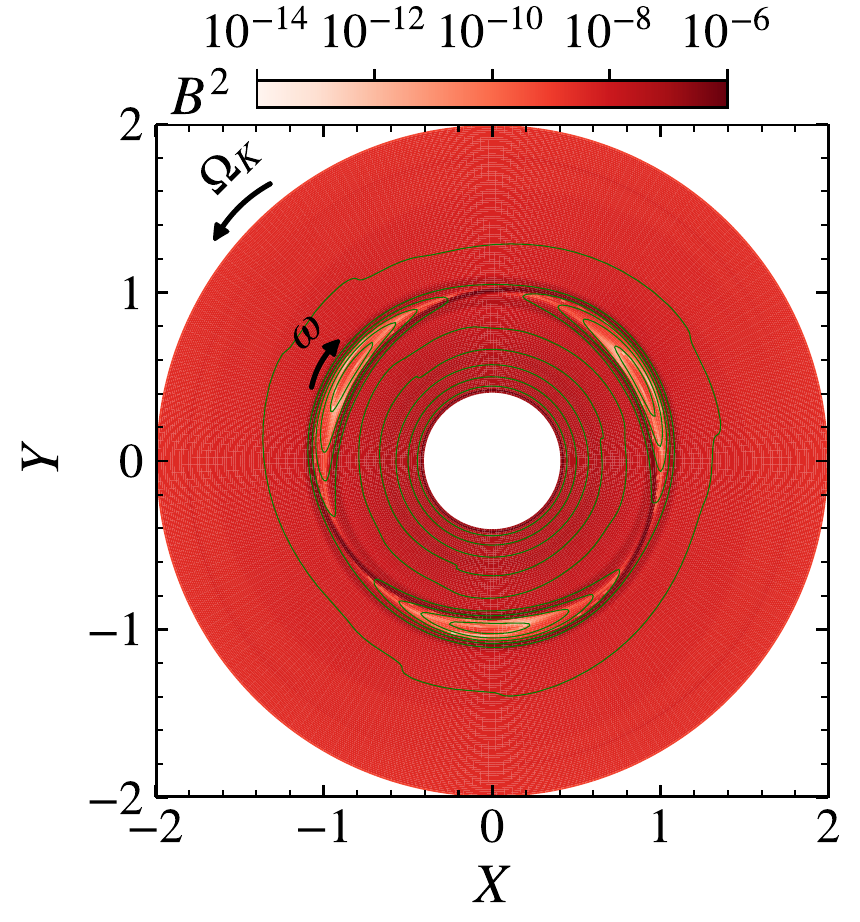}
        \caption{Plot of $B^2$ at late times ($t=30\approx5P_0$) for $\beta=5\times10^5$  across the entire simulated disc. Here $(X,Y) = (R\cos\phi, R\sin\phi)$ are the Cartesian coordinates associated with the disc. Magnetic field strength is shown in monochromatic red-scale and contours of $\Sigma$ are plotted in green to indicate the location of the vortices. Along the edges of the vortices, the field has grown by nearly a factor of 10 from its initial value and the magnetic pressure is roughly equipartitioned between the toroidal and radial components.}
\label{fig:B2_late_time}
\end{figure}

\begin{figure}
                \centering
                \includegraphics[width=0.45\textwidth]{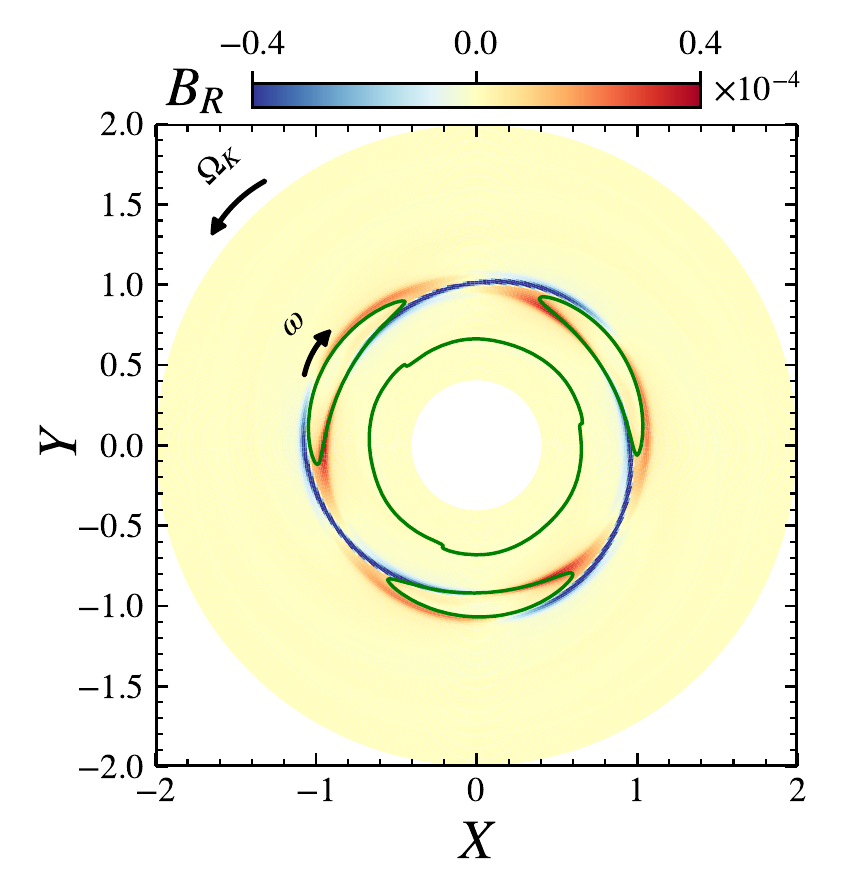}
        \caption{Plot of radial magnetic field $B_R$ at late times ($t=30\approx5P_0$) for  $\beta=5\times10^5$  across the  entire simulated disc. Here $(X,Y) = (R\cos\phi, R\sin\phi)$ are the Cartesian coordinates associated with the disc. Red indicates positive radial field, blue indicates negative radial field and yellow indicates zero radial field. A single contour $\Sigma = 1.5$ (green) indicates the location of the vortices at $t = 30$. The radial field as a whole follows a pattern predicted by the kinematic model of Section \ref{sec:field_twisting}, whereby the leading edge is a radial field sink and the trailing edge a radial field source.}
\label{fig:Br_late_time}
\end{figure}

At late times, the magnetic flux has diffused out of the vortex and concentrates around its edges. In Fig. \ref{fig:B2_late_time}, we plot the magnetic pressure $P_B\sim B^2$ for $t=30\approx5P_0$ for our lowest-magnetisation case. The magnetic field has grown by nearly a factor of 10 along the edges of the vortices and decreased nearly to zero inside the vortices. The magnetic pressure is roughly equipartitioned between the toroidal and radial components of the field, indicating highly twisted field lines. 

The rotation of the vortex generates a radially directed field at its lobes. The disc is orbiting in the anticlockwise direction so the anticyclonic vortex is rotating clockwise. The initially toroidal field wraps around disc centre in the anticlockwise direction. The configuration as a whole is akin to Moffatt's kinematic model. We thus expect the leading edge of the vortex to act like a sink of radial magnetic field and the trailing edge like a source. In Fig. \ref{fig:Br_late_time}, we show this characteristic pattern in the $B_R$ field for our simulations. We indicate the direction of rotation of the disc and the vortices to make more explicit the correspondence between the radial field direction in our simulations and our model shown in Fig. \ref{fig:late_time}.

\section{Discussion}
\label{sec:discussion}
We have investigated the RWI in the context of an accretion disc threaded by a poloidal magnetic field. We summarise our main findings below. 

\smallskip

\noindent{\bf 1.} For weak magnetic fields ($\beta > 10^3$), vortex growth rates correspond to those of the reference HD simulation. For stronger fields ($10^1 \leq \beta \leq 10^3$), the growth rate scales like $\gamma \sim \beta^{0.32} \ \Omega_K$. Hence, toroidal magnetic fields suppress the vortex growth rate.

\smallskip

\noindent{\bf 2.} Vortex growth halts after a few orbits and density saturates to a late-time value. For $\beta > 10^3$, this amplitude corresponds to the reference HD saturation value. For $10^2 \leq \beta \leq 10^3$, the saturation value scales like $\psi_{\infty} \sim \beta^{0.075}\psi_0$. Hence, toroidal magnetic fields suppress the late-time amplitude of the perturbation.

\smallskip

\noindent{\bf 3.} At early times, before turbulent diffusion takes over, the radial field grows linearly. The field becomes twisted, converting toroidal field into radial field near the edges of the vortex. Because the vortices are anti-cyclonic, a toroidal field pointing in the direction of the rotation of the disc produces a positive $B_R$ near the leading edge of the vortex and a negative $B_R$ along the trailing edge.

\smallskip 

\noindent{\bf 4.} At late times, after the turbulent diffusion dominates, the magnetic flux is expelled from the vortex and onto its boundary. The resulting magnetic field has a characteristic pattern where the leading edge acts as a sink of $B_R$ field and the trailing edge acts as a source. 

\smallskip

The RWI provides a mechanism for trapping dust in protostellar discs, which can serve as a seed for planetesimal formation. We find that at late times, the magnetic field assumes a characteristic pattern outside the vortex region. This may affect the trapping of dust in the vortex and will be the subject of future investigation. 

Previous work has shown that the RWI is primarily a 2D instability. However, in the context of understanding the evolution of the magnetic field, the addition of vertical disc structure may be important. 3D simulations of the vortex may allow the formation of a vertical component of the magnetic field from the toroidal component. The late-time field could then be characterised via a multipole expansion to determine which modes preferentially form outside the vortex. This additional field component may also affect dust trapping in 3D, which is already expected to be more complex due to vertical dust circulation.

The presence of a magnetic field can also be interesting in the context of vortex mergers. If the time scale for vortices merging is long compared to the turbulent diffusion time, our fourth finding above shows that magnetic flux would be concentrated along the vortex edges when mergers occur. Vortices adjacent to each other in the disc would have opposite polarity. When they merge, the flux would annihilate, heating the disc and possibly leading to thermally driven outflows. If these outflows are faster than the escape velocity of the system, they may be observable with ALMA analogously to protostellar outflows.

Finally, a possible method of detecting magnetic fields in protoplanetary discs is polarised IR emission. Thermally emitting grains are thought to align with the magnetic field (Lazarian 2007). Polarization measurements of HL Tau with CARMA by Stephens et al. (2014) found evidence of a toroidal field, but recent ALMA observations question if the morphology of polarised emission is due to magnetic fields (Stephens et al. 2017). A coronagraphic instrument on WFIRST will also provide polarimetric measurements of protoplanetary discs. This motivates the need for self-consistent disc models that include both dust-trapping and magnetic fields, in order to answer the question of whether polarised emission is due to dust tracing the magnetic field.

\section*{Acknowledgments}
This work was supported in part by NASA grants NNX12AI85G and NNX14AP30G and by NSF grant AST 1211318. L. Matilsky was supported during this work by a NY NASA Space Grant Fellowship at Cornell University during the summer of 2014 and by the George Ellery Hale Graduate Fellowship at the University of Colorado Boulder during the summer of 2018. L. Matilsky thanks his advisors Juri Toomre and Bradley W. Hindman at JILA, University of Colorado Boulder, for their support and encouragement during the final revisions of the manuscript in the summer of 2018. We thank the \textsc{PLUTO} team for allowing us to use their excellent MHD code, without which this work would not have been possible. We thank University of Colorado Boulder graduate student Nicole Arulanantham  for her helpful tip on using the gas-rich disc surrounding V4046 Sgr for comparison to the non-dimensional discs in our models. We thank the anonymous referee for helpful feedback in revising this manuscript, which, we believe, is of significantly greater quality than the originally submitted version as a result.


\label{lastpage}

\end{document}